\documentclass[%
 superscriptaddress,
 reprint,
 showpacs,preprintnumbers,
 nofootinbib,
 amsmath,amssymb,
 aps,
 prd,
 floatfix,
]{revtex4-2}
\usepackage{graphicx}
\usepackage{xcolor}
\usepackage{array,url}
\usepackage{numprint}
\usepackage{hyperref}
\hypersetup{
    colorlinks=true,
    linkcolor=blue,
    filecolor=blue,
    urlcolor=blue,
    pdftitle={Longitudinal\_kinematic\_imbalances},
    }

\usepackage{amssymb,amsmath,amstext,amsthm,amsfonts,nicefrac}
\usepackage[capitalise]{cleveref}

\usepackage{tabularx, makecell}
\usepackage{booktabs}
\usepackage{xspace}
\usepackage{xstring}

\usepackage{hyperref}
\usepackage{ulem}
\usepackage{multirow}
\usepackage{amsmath}
\usepackage{accents}
\usepackage{xcolor,colortbl}

\usepackage[range-units=single]{siunitx}

\DeclareSIUnit\eV{\electronvolt}
\DeclareSIUnit\eVperc{eV/\text{\ensuremath{c}}}
\DeclareSIUnit\pot{POT}

\newcommand{\deltacp}{\delta_\textrm{CP}}

\newcommand{\pt}{p_\textrm{T}}
\newcommand{\plong}{p_\textrm{L}}
\newcommand{\plrec}{p_\textrm{L,rec}}

\newcommand{\dpt}{\delta \pt}

\newcommand\pbar{\smash[b]{\raisebox{0.33\height}{\scalebox{0.6}{\tiny\textbf{(}\,}}{\mkern-1.5mu\textbf{---}\mkern-1.5mu}\raisebox{0.33\height}{\scalebox{0.6}{\,\tiny\textbf{)}}}}}
\newcommand*{\parbar}[1]{\accentset{\pbar}{#1}}

\newcommand{\pnl}{p_{N,\textrm{L}}}
\newcommand{\pni}{p_{N,i}}
\newcommand{\ppl}{p_{N^\prime,\textrm{L}}}
\newcommand{\pmul}{p_{\mu,\textrm{L}}}

\newcommand{\ermv}{E_\textrm{rmv}}

\newcommand{\dplvis}{\delta p_\textrm{L,vis}}
\newcommand{\evis}{E_\textrm{vis}}

\newcommand{\dpty}{\delta p_{\textrm{T}y}}
\newcommand{\emnu}{E_{\textrm{m},\nu}}
\newcommand{\emnutrue}{E_{\textrm{m},\nu,\textrm{true}}}

\newcommand{\fa}{F_\textrm{A}(Q^2)}

\newcommand{\eq}[1]{\begin{align}#1\end{align}}

\newcommand{\ccite}[1]{%
\IfSubStr{#1}{,}{Refs.~}{Ref.~}\cite{#1}%
}

\newcommand{\quotes}[1]{``#1''}

\newcommand{\ie}{i.e.\@\xspace}

\RequirePackage{xspace}

\newcommand{\sfgd}{Super-FGD\xspace}

\newcommand{\hk}{HK\xspace}
\newcommand{\minerva}{MINERvA\xspace}

\newcommand{\affiliationETHZ}{ETH Zurich, Institute for Particle physics and Astrophysics, CH-8093 Zurich, Switzerland.}
\newcommand{\affiliationCERN}{European Organization for Nuclear Research (CERN), 1211 Geneva 23, Switzerland.}

\newcommand{\affiliationCEA}{IRFU, CEA, Université Paris-Saclay, F-91191 Gif-sur-Yvette, France.}

\begin{document}

	\title{Longitudinal kinematic imbalances in neutrino and antineutrino interactions for improved measurements of neutrino energy and the axial vector form factor}

	\author{N.~Baudis}
	\email[Contact e-mail: ]{nathan@baud.is}
	\affiliation{\affiliationETHZ}

	\author{S.~Dolan}
	\email[Contact e-mail: ]{Stephen.Joseph.Dolan@cern.ch}
	\affiliation{\affiliationCERN}

	\author{D.~Sgalaberna}
    \email[Contact e-mail: ]{davide.sgalaberna@cern.ch}
    \affiliation{\affiliationETHZ}

	\author{S.~Bolognesi}
	\affiliation{\affiliationCEA}

	\author{L.~Munteanu}
	\affiliation{\affiliationCERN}

    \author{T.~Dieminger}
	\affiliation{\affiliationETHZ}

\begin{abstract}
Current and future accelerator neutrino oscillation experiments require an improved understanding of nuclear effects in neutrino-nucleus interactions.
One important systematic uncertainty is introduced by the collective impact of nuclear effects which bias the reconstruction of the neutrino energy, such as the nuclear removal energy.
In this manuscript, we introduce a novel observable for accelerator neutrino oscillation experiments, the visible longitudinal momentum imbalance, reconstructed in charged current quasi-elastic interactions from the outgoing charged lepton and nucleon. 
We demonstrate it to be minimally dependent on the neutrino energy and sensitive to sources of bias in neutrino energy reconstruction.
Furthermore, we show how the use of the longitudinal imbalance in antineutrino interactions in a target containing hydrogen allows for an improved, high-purity selection of the interactions on hydrogen. This approach offers the potential for precise measurements of the nuclear axial vector form factor as well as of the antineutrino flux.
\end{abstract}

\maketitle

\section{Introduction}
\label{sec:introduction}

Long-baseline (LBL) neutrino oscillation experiments have been delivering increasingly precise measurements of the neutrino oscillation parameters \cite{T2K:2021xwb,T2K:2023smv,NOvA:2021nfi}, in particular towards the determination of the leptonic CP-violating phase $\deltacp$.
With increasing precision, a reduction of the systematic uncertainties associated with the modeling of neutrino interactions with nuclear targets is vital~\cite{Alvarez-Ruso:2017oui}. 
To this end, near detectors close to the neutrino production site are employed to constrain the flux before oscillation as well as the neutrino-nucleus interaction model. 
Tokai-to-Kamiokande (T2K)~\cite{Abe:2011ks} is an LBL experiment located in Japan measuring the oscillation of a predominantly muon (anti)neutrino beam into muon and electron neutrinos over a baseline of \SI{295}{\kilo\meter}. The future Hyper-Kamiokande (\hk) experiment~\cite{Hyper-Kamiokande:2018ofw} will utilize the same baseline and near detector suite, with a far detector increased in size.
At the T2K/\hk neutrino beam energy with a peak at \SI{600}{\mega\eV}, the dominant neutrino interactions are charged-current quasi-elastic (CCQE) interactions such as $\nu_\mu + n \rightarrow \mu^- + p$, which occur on a single nucleon within the target nucleus within the impulse approximation.
CCQE interactions on nuclear targets are obfuscated by both the final state interactions (FSI) of the outgoing nucleon with the nucleus, which can change the final state particle kinematics and content of the interactions, as well as by the initial \quotes{Fermi} motion of nucleons and the nuclear removal energy required to liberate them. The distribution and correlation of initial state nucleon momentum and removal energy can be described by a so-called spectral function.
Due to the broad neutrino flux spectrum, the energy of the incoming neutrino is unknown on an event-by-event basis, complicating any attempt to constrain nuclear effects, which in turn bias the neutrino energy reconstruction that is relied upon in oscillation measurements. However, the knowledge of the beam direction can be exploited by using kinematic imbalances in the transverse plane. Such single transverse variables (STVs) have been extensively studied and shown to offer constraints on nuclear effects such as the Fermi motion, multi-nucleon correlations and FSI~\cite{Lu:2015tcr,Dolan:2018sbb,Dolan:2018zye,Ershova:2022sof}.
The ongoing upgrade of the T2K off-axis near detector is well equipped to measure such imbalances, where the fully active, 3D segmented plastic scintillator detector \sfgd~\cite{Blondel:2017orl, Blondel:2020hml, Abe:2019whr} both improves the proton detection threshold and is capable of reconstructing the momenta of outgoing neutrons~\cite{Dolan:2022sut,Abe:2019whr,Munteanu:2019llq}.
The latter enables the measurement of STVs in antineutrino interactions, shown to isolate interactions on hydrogen, free of nuclear effects~\cite{Munteanu:2019llq}.
However, STVs offer limited power to constrain effects which shift the overall final state energy in relation to the unknown neutrino energy, such as the nuclear removal energy~\cite{Dolan:2022sut,Bodek:2018lmc}.
A mismodeling of the removal energy biases the reconstructed neutrino energy and in turn the measurement of the neutrino oscillation parameters.
This affects the measurement of the mass difference squared $\Delta m^2_{23}$ in particular, which is directly related to the neutrino energy at a given baseline, but also can form a primary systematic error for a measurement of $\deltacp$~\cite{T2K:2019bcf}.
With the removal energy forming a major systematic uncertainty in neutrino oscillation measurements at T2K, recent analyses have included more sophisticated modeling of its associated uncertainties~\cite{T2K:2021xwb,T2K:2023smv, Chakrani:2023htw}. Whilst these uncertainties can in principle be constrained from precision electron scattering measurements, the precondition for this is that observations can be interpreted in terms of intrinsic nuclear ground state properties, independent from the interaction probe. The widely-used factorization ansatz~\cite{Benhar:2005dj,Benhar:2006wy} permits this at intermediate to large momentum transfers in which the impulse approximation is well known to hold but, beyond this, some model-dependent corrections are required~\cite{Lapikas:1993uwd,Ankowski:2014yfa}. 
Furthermore, no neutrino interaction model (extracted from electron scattering measurements or otherwise) has been shown to produce satisfactory agreement with global neutrino scattering data~\cite{Avanzini:2021qlx}. With this in mind, in-situ neutrino scattering measurements which are sensitive to the leading systematic uncertainties for neutrino oscillation analyses are of crucial importance to ensure that oscillation measurements are both precise and robust, as evidenced by the well-established aforementioned utility of STV measurements.
In this paper, we introduce a novel observable characterizing the longitudinal kinematic imbalance which is directly sensitive to nuclear effects which cause bias in neutrino energy reconstruction and is minimally dependent on the neutrino energy. The observable can be measured at near detectors of neutrino oscillation experiments to benchmark input models and constrain uncertainties. Further, we show how it may be employed in antineutrino interactions to obtain a high-purity sample of interactions on hydrogen nuclei within a composite nuclear target.

\section{Longitudinal kinematic imbalance}

Consider an (anti)neutrino CCQE interaction in the impulse approximation, occurring with a neutron (proton) $N$ within a nucleus with $A$ nucleons and producing a proton (neutron) $N^\prime$ in the final state: $\parbar{\nu}_\mu + N \rightarrow \mu^\mp + N^\prime$. 
The energy and momentum conservation read, where the latter is split in the transverse (T) and longitudinal (L) directions:
\eq{
p_\nu + \pnl &\approx \overbrace{\pmul + \ppl}^{\plong} \label{eq:ccqe_ia_pl},\\
\vec{p}_{N,\textrm{T}} &\approx \vec{p}_{\mu,\textrm{T}} + \vec{p}_{N^\prime,\textrm{T}} \label{eq:ccqe_ia_pt},\\
E_\nu &\approx \underbrace{E_\mu + E_{N^\prime} - m_N}_{\evis} + \ermv. \label{eq:ccqe_ia_e}
}
The quantities on the right-hand side refer to the final state energies and momenta, where the equalities are inexact due to final state interactions of the outgoing particles with the nuclei's strong and Coulomb potentials \cite{Furmanski:2016wqo,Hayato:2021heg,Bodek:2018lmc}. 
While negligible in certain cases~\cite{Cai:2020nbe} but not in general~\cite{Bodek:2018lmc}, their effect is discussed further below.
The removal energy $\ermv$ denotes the contribution to the neutrino energy which is undetected due to nuclear effects, where the equation above holds pre-FSI: $\ermv = E_\nu - \evis^\textrm{pre-FSI}$. It is given by the sum of both the nuclear excitation and separation energies, as well as the small kinetic energy of the nuclear remnant: $\ermv = E_x + S^N + T_{A-1}$. The separation energy for a nucleon $N$ reads $S^N = M_{A-1} + M_N - M_A$, while the excitation energy is given by $E_x = M_{A-1}^* - M_{A-1}$. Here, $M_A$ denotes the mass of the initial nucleus, while $M_{A-1}^*$ and $M_{A-1}$ are the mass of the excited and de-excited remnant nucleus, respectively \cite{Bodek:2018lmc,Cai:2020nbe}.
The transverse momentum imbalance reads $\dpt \equiv |\vec{p}_{\mu,\textrm{T}} + \vec{p}_{N^\prime,\textrm{T}}|$~\cite{Lu:2015tcr}.
We introduce the visible longitudinal momentum imbalance as an observable given by:
\eq{\label{eq:dplvis_def}
\dplvis \equiv \plong - \evis/c,
}
where $\plong$ and $\evis$ are the overall final state longitudinal momentum and visible energy, respectively, as indicated above.
From \cref{eq:ccqe_ia_pl,eq:ccqe_ia_e}, $\dplvis$ yields
\eq{\label{eq:dplvis_pnl_ermv}
\dplvis \approx \pnl + \ermv/c.
}
In the absence of nuclear effects, such as for neutrino interactions on a hydrogen target, $\dplvis$ is thus zero. For CCQE interactions in more complex nuclei it becomes sensitive to the nucleon initial longitudinal momentum and nuclear removal energy, where in the equation above the latter contributes on the order of \SI{15}{\percent} in magnitude for common target nuclei such as carbon, oxygen and argon~\cite{Bodek:2018lmc}.
For a given event, the two can only be detected in sum.
It should be noted that $\dplvis$ is similar in concept to the reconstructed longitudinal nucleon momentum $\plrec$ proposed in \ccite{Furmanski:2016wqo}.  
However, in the absence of strong and Coulomb potentials, it is directly sensitive to the removal energy, as no assumption about its distribution is made, unlike when computing $\plrec$.
Note also that, whilst $\dplvis$ is introduced above in the context of CCQE interactions within the impulse approximation, it can be easily generalized to other interaction channels and final state topologies by extending the list of final state particles considered in the calculation of $\plong$ and $\evis$. In general the impact of FSI considered via nuclear cascades in neutrino event generators~\cite{Dytman:2021ohr} can be expected to distort $\dplvis$ but 
leave the correlation between measurements of a post-FSI $\dplvis$ and removal energy intact.
Further, more sophisticated treatments of FSI in modern microscopic models (see e.g.~\cite{Nikolakopoulos:2022qkq,Franco-Patino:2022tvv}) likewise affects the visible final state energy in relation to the true neutrino energy~\cite{Ankowski:2014yfa} via a consideration of the impact of the nuclear potential on the outgoing nucleon, creating additional potential that must be overcome (and so effectively adding additional terms to the right hand side of \cref{eq:ccqe_ia_e} and \cref{eq:dplvis_pnl_ermv}). We therefore expect $\dplvis$ to provide sensitivity to the overall \quotes{missing} neutrino energy defined simply as:
\begin{equation}\label{eq:em_nu}
    \emnu \equiv E_\nu - \evis.
\end{equation}

This includes the collective impact of the nuclear potentials on the bias in neutrino energy reconstruction, including both removal energy and nuclear potential effects (in addition to small Coulomb corrections), which as mentioned is the source of a major uncertainty in neutrino oscillation experiments.

\begin{figure}[tb]
    \centering
    \includegraphics[width=\linewidth]{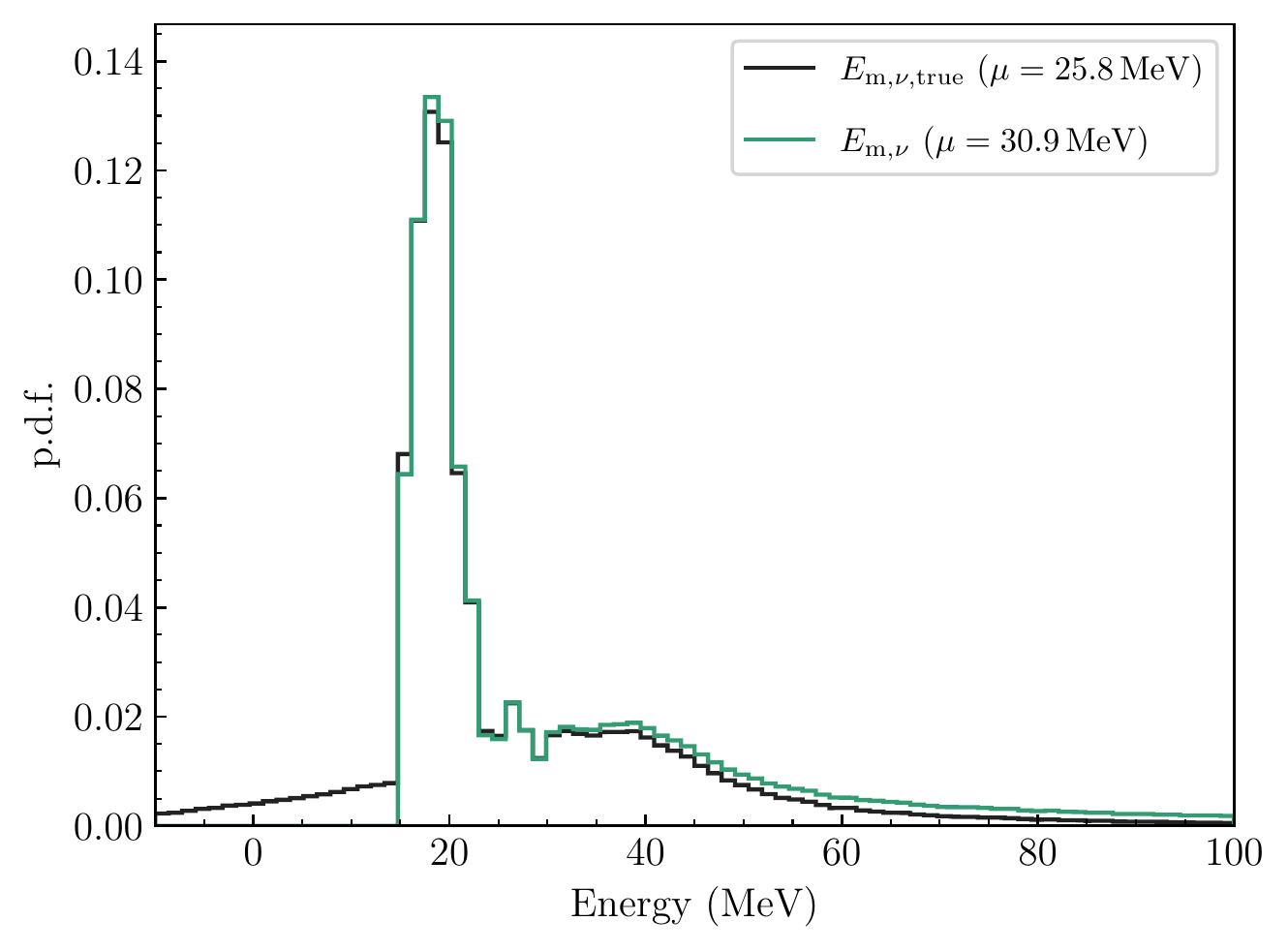}
    \caption{The NEUT prediction for the missing neutrino energy $\emnu$ in neutrino CCQE interactions including FSI, as well as the \quotes{true} missing energy $\emnutrue$ reconstructed from all outgoing particles. As for all subsequent figures, the T2K neutrino flux is used as input with polystyrene-based scintillator (C$_8$H$_8$) as the target material.}
    \label{fig:emiss_nu}
\end{figure}

\begin{figure}[tb]
    \centering
    \includegraphics[width=\linewidth]{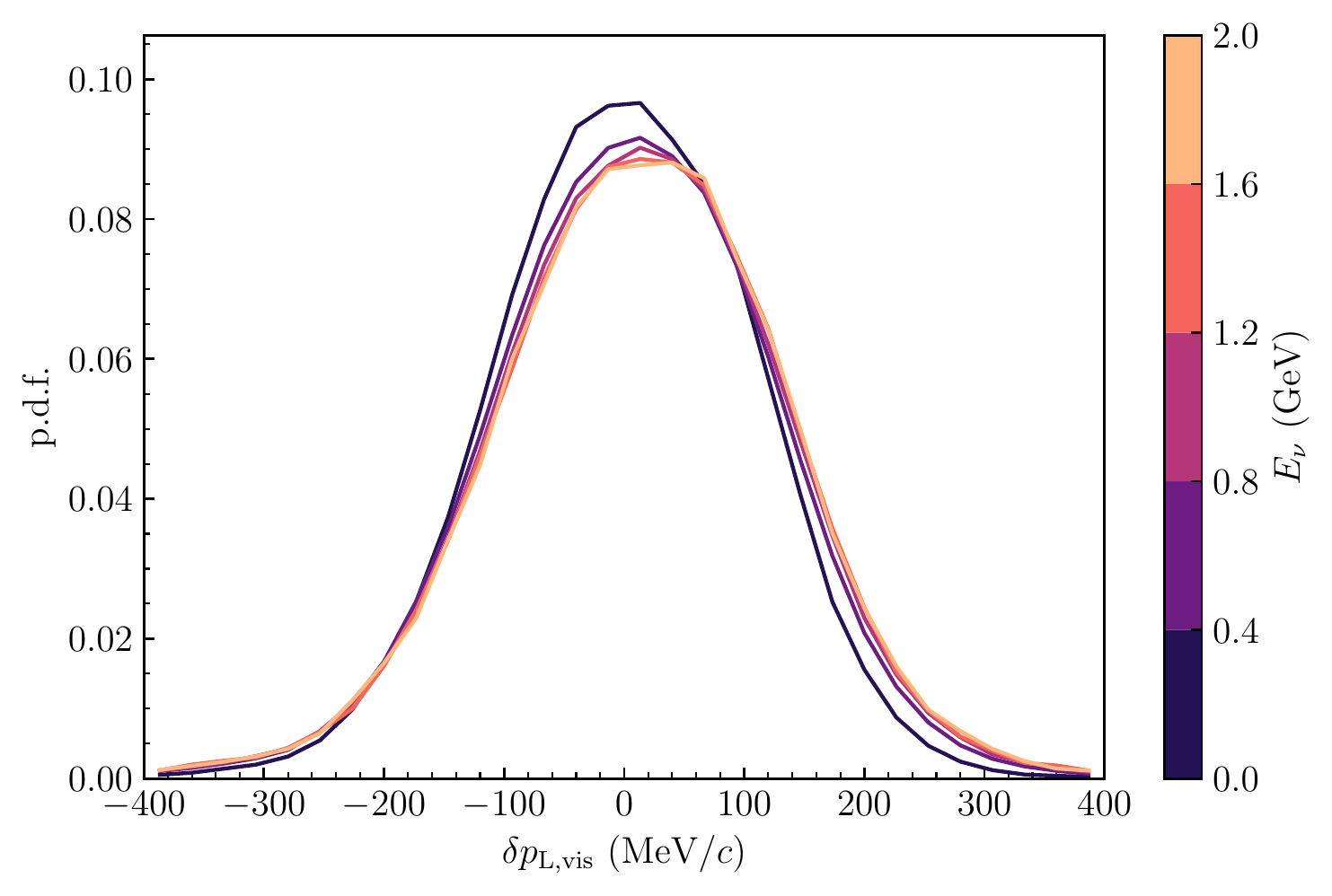}
    \caption{The NEUT prediction for $\dplvis$ in CCQE neutrino interactions on polystyrene-based scintillator (C$_8$H$_8$) in bins of the neutrino energy, indicated on the right. The T2K flux was used as input, where for events undergoing FSI, all particles created by FSI are considered.}
    \label{fig:dplvis_etrue_bins}
\end{figure}

We evaluate the properties of $\dplvis$ within a modern simualtion using the NEUT neutrino interaction generator~\cite{Hayato:2021heg}, which is used as an input to the T2K experiment's neutrino oscillation analyses~\cite{T2K:2023smv}, considering the flux of muon neutrinos the experiment expects to observe at its ND280 near detector~\cite{T2K:2012bge,t2kfluxurl}. NEUT describes  the initial state using the Spectral Function (SF) model from Ref.~\cite{Benhar:1994hw}. The SF model employs the plane wave impulse approximation to apply the factorize ansatz, thereby separating  the pre-FSI fully exclusive CCQE cross section into terms in which all nuclear dynamics are encoded within a two dimensional spectral function relating removal energy to initial state nucleon momentum, which is extracted from exclusive electron scattering measurements. In general, the projection of the spectral function onto the removal energy axis results in sharp spike corresponding to shell-model states on top of a background related to nucleon correlation effects~\cite{Lapikas:1993uwd,Benhar:1994hw}. Refs.~\cite{Chakrani:2023htw,Hayato:2021heg, T2K:2023smv} provide figures showing this function. Within NEUT, FSI is modelled by propagating simulated nucleons through an intranuclear cascade which both alters their kinematics and predicts the emission of additional hadrons~\cite{Hayato:2021heg,Dytman:2021ohr} but does not consider the FSI modification to the inclusive cross-section that would be captured using a microscopic description of the nuclear potential.

In \cref{fig:emiss_nu}, we show the missing neutrino energy as defined in \cref{eq:em_nu} for interactions generated with the NEUT SF model, including FSI. Here, $\evis$ is computed using the outgoing muon and leading (highest-momentum) proton. In addition, we show a \quotes{true} missing neutrino energy, obtained from reconstructing $\evis$ using all particles created in the FSI cascade, which is then equivalent to removal energy.
The missing neutrino energy reconstructed from the leading proton closely follows the shape of the underlying removal energy distribution, described above, while including all particles shifts and smears out the energy deficit, extending to negative values.

\begin{figure}[tb]
    \centering
    \includegraphics[width=\linewidth]{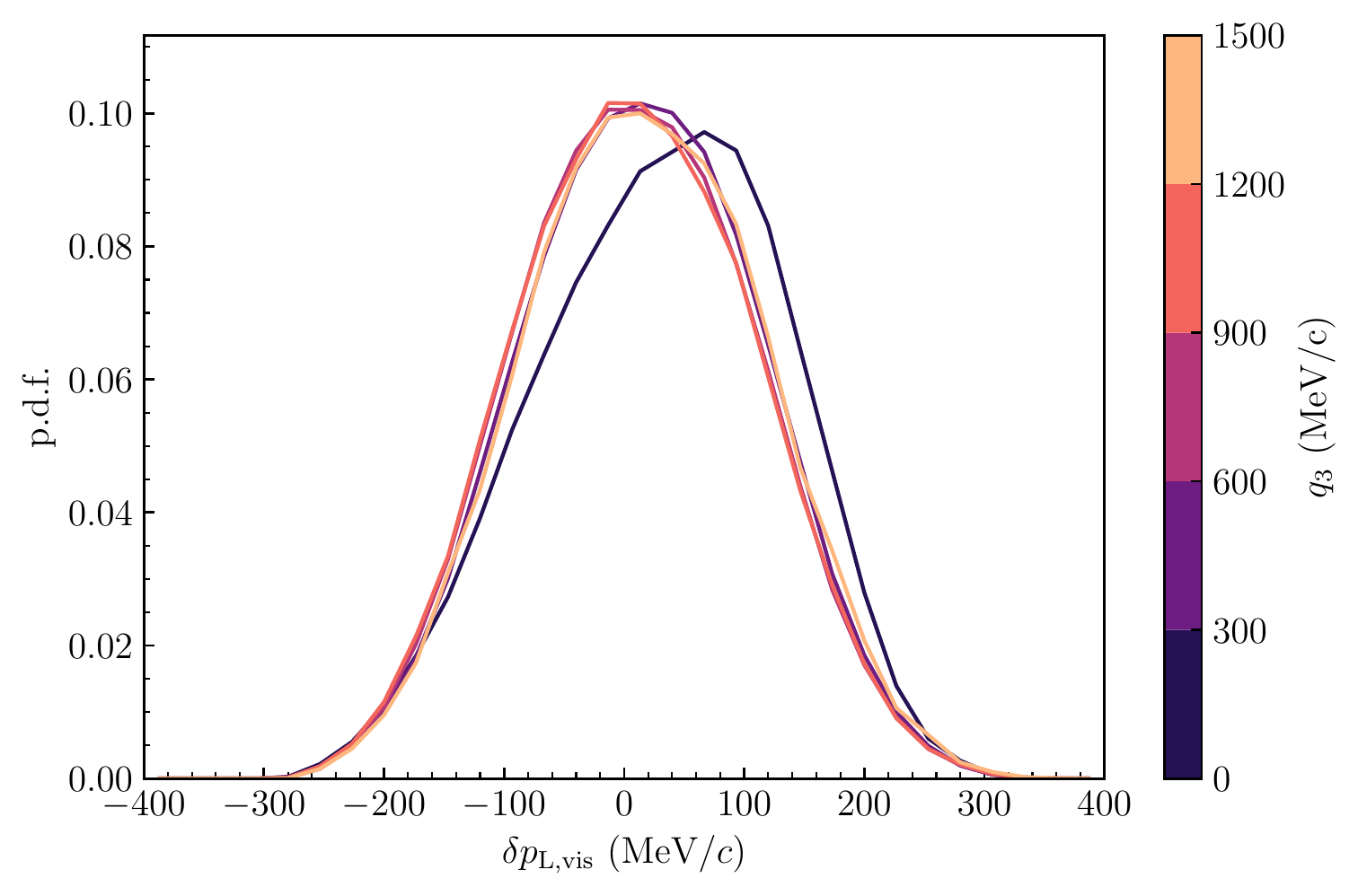}
    \includegraphics[width=\linewidth]{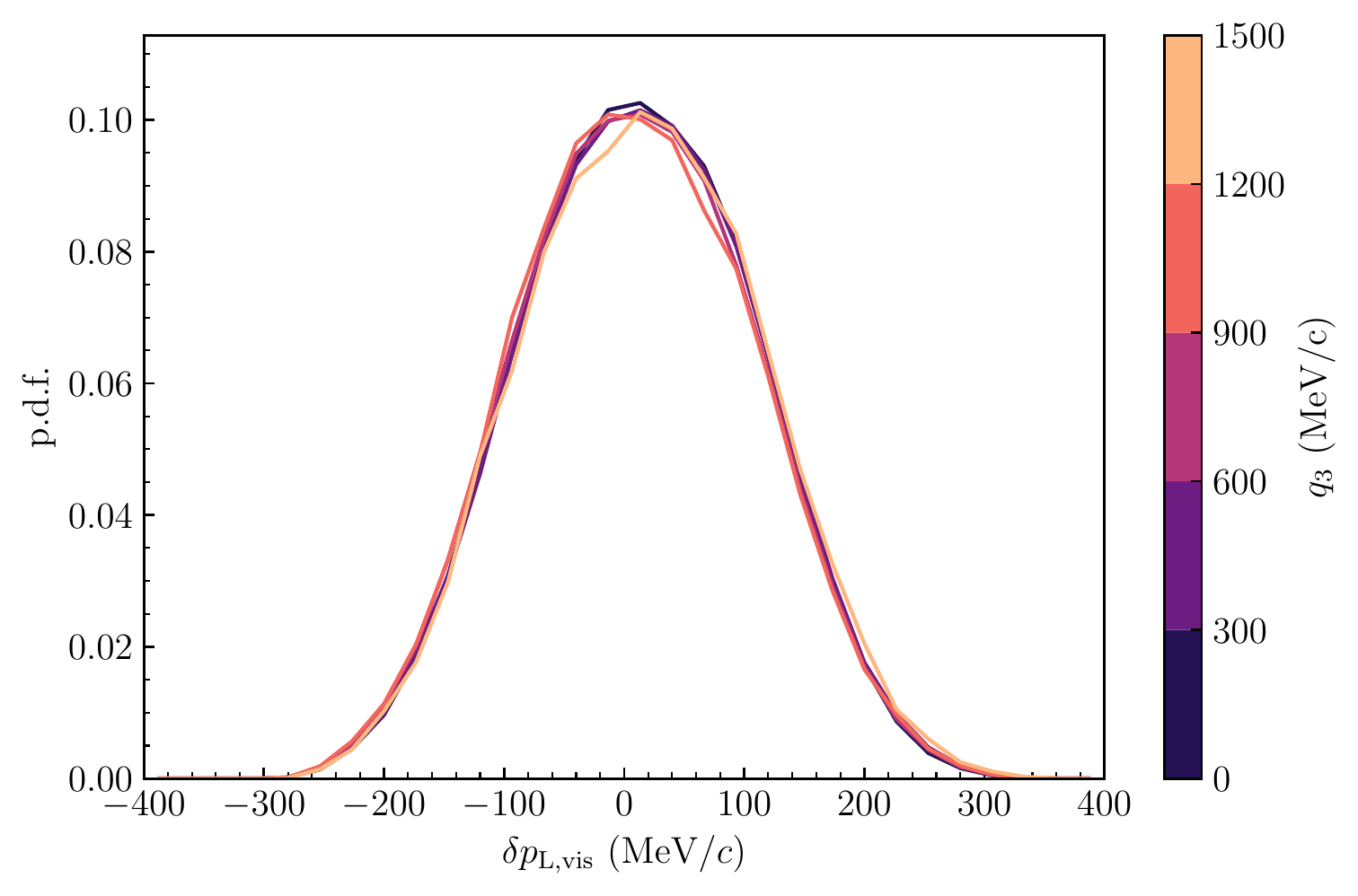}
    \caption{The $\dplvis$ distributions in neutrino CCQE interactions without FSI in bins of the magnitude of momentum transfer $q_3$, with (top) and without (bottom) Pauli blocking.
    }
    \label{fig:dpLvis_pauli_blocking}
\end{figure}

As demonstrated above, we expect $\dplvis$ to be independent of the neutrino energy to first order. 
We confront this expectation with the model prediction from NEUT in \cref{fig:dplvis_etrue_bins}, where a minimal dependence on the neutrino energy can indeed be observed, in particular at energies above \SI{0.4}{\giga\eV}.
Similarly to the transverse momentum imbalance~\cite{Lu:2015tcr}, a small dependence remains due to second-order effects such as Pauli blocking.
While the underlying distribution of the initial nucleon longitudinal momentum $\pnl$ is isotropic, nucleons with a momentum opposite that of the neutrino have an increased interaction cross section due to the higher center of mass energy, causing the $\pnl$ distribution sampled by neutrino interaction to be biased towards negative values, creating a polarization effect. 
Pauli blocking on the other hand will cause a positive bias in the initial longitudinal momentum.
With the momentum transfer to the hadronic state primarily occurring along the neutrino direction, initial nucleons with a forward momentum are less likely to undergo Pauli blocking, where the magnitude of this effect decreases at higher neutrino energies. 
This is illustrated in \cref{fig:dpLvis_pauli_blocking}, which shows $\dplvis$ distributions in CCQE interactions binned in the magnitude of momentum transfer $q_3$. It can be seen how at low momentum transfers, there is a positive bias in the $\pnl$ distribution due to Pauli blocking, where nuclei with lower momenta are more likely to inhabit an occupied state after the interaction.
Without Pauli blocking, the $\dplvis$ distribution is seen to be largely independent of the momentum transfer.
The net shift on $\dplvis$ from these effects is predicted to be on the level of \SIrange{2}{3}{\mega\eVperc} for the T2K~flux.
Here we note that when analyzing the distribution of $\pnl$ in CCQE interactions, we observed an unexplained shift in the NEUT output for the SF model of around \SI{-10}{\mega\eVperc}; details are reported in Appendix~\ref{sect:appendix_sf_shift}.

As the Fermi motion is isotropic, the distribution of the initial longitudinal momentum $\pnl$ is identical to the distribution of the two transverse components. Accounting for the second-order effects mentioned above, the observed transverse momentum imbalance can thus be used to gain information on the $\pnl$ distribution.
Using \cref{eq:dplvis_pnl_ermv}, this constraint can be propagated to the observed $\dplvis$ distribution to statistically obtain constraints on the missing neutrino energy as is explained in detail below.

\section{Analysis}
\label{sec:analysis}

\subsection{Sensitivity to missing neutrino energy} %

\begin{figure}[htb]
    \centering
    \includegraphics[width=\linewidth]{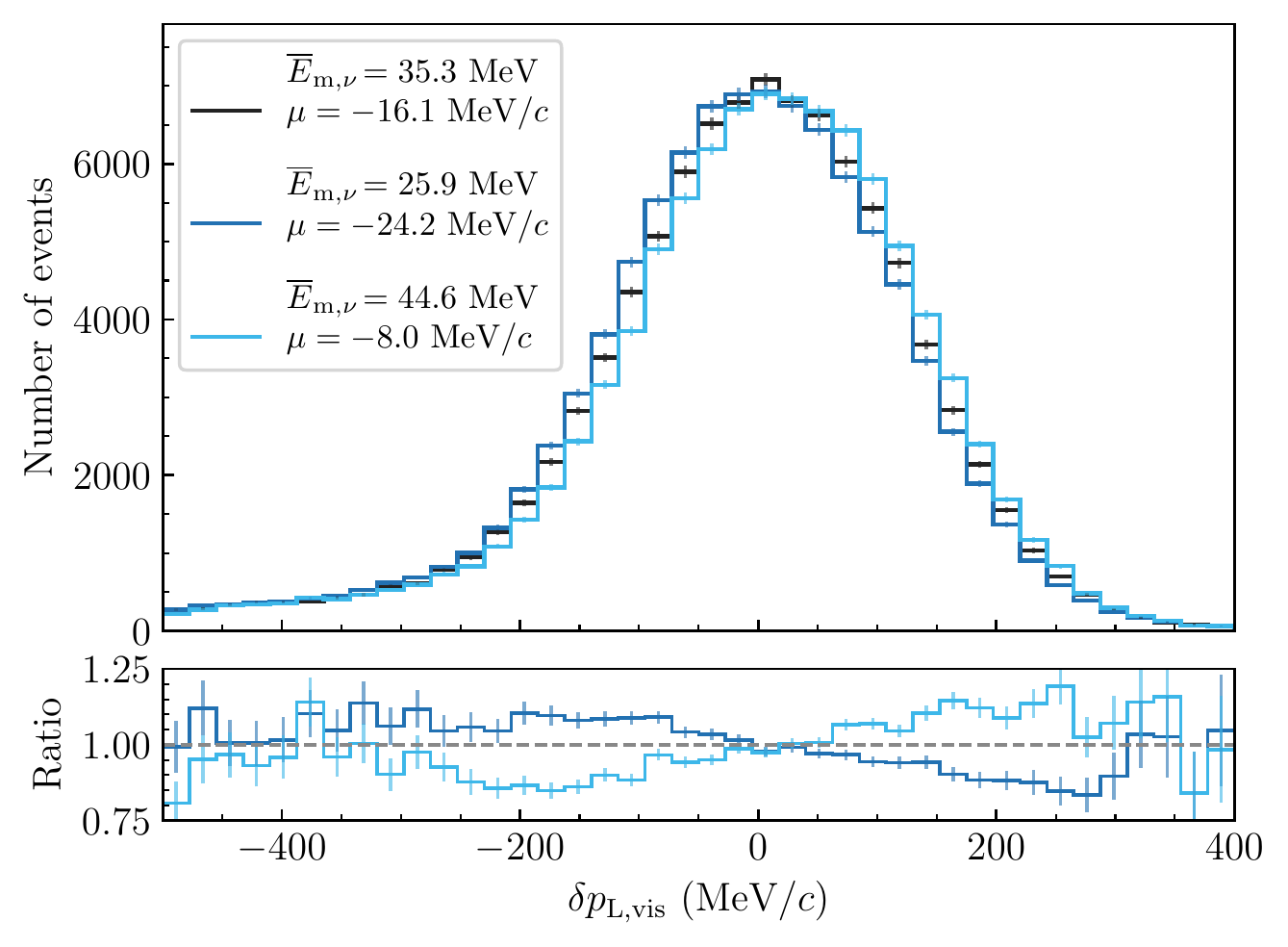}
    \caption{The $\dplvis$ distributions for the nominal NEUT SF model and the same model with the removal energy distribution shifted downwards and upwards by \SI{10}{\mega\eV}. The corresponding shifts in the average missing neutrino energy of the CCQE component are indicated in the legend ($\overline{E}_{\textrm{m},\nu}$), alongside the means of the overall $\dplvis$ distributions ($\mu$). The T2K neutrino flux is used for \SI{4e21}{\pot}, where detector
    smearing is applied to CC0$\pi$ events with one muon and one proton detected in the final state. The statistical errors are indicated, while the lower panel shows the ratio of the shifted models to the nominal model in each bin. %
    }
    \label{fig:dplvis_sf_models_4e21pot}
\end{figure}

We proceed here with an analysis to illustrate how measuring~$\dplvis$ in neutrino interactions may be used to constrain the nuclear effects which cause bias in neutrino energy reconstruction. %
We consider a selection of neutrino interactions generated with NEUT and apply a detector smearing analogous to the one of the \sfgd, 
using parametrized detector efficiencies and resolutions as in \ccite{Dolan:2022sut,Munteanu:2019llq}.
Note again that NEUT does not directly consider the impact of the nuclear potential. However, by example of this model and the sensitivity of $\dplvis$ to the removal energy within it we show sensitivity of the observable to effects which bias the neutrino energy reconstruction.
In particular, we show how the overall distribution of $\dplvis$ can deliver information on both the average removal energy as well as the shape of its distribution.
Firstly, the mean of $\dplvis$ depends on the average removal energy. This effect is shown in \cref{fig:dplvis_sf_models_4e21pot}, where the $\dplvis$ distribution of the nominal NEUT SF model is compared to the same models with removal energy distributions shifted by $\pm \SI{10}{\mega\eV}$. The corresponding shifts in the average missing neutrino energy are indicated. We include the effects of 
detector smearing and a CCQE-like selection for pion-less (CC0$\pi$) topologies, which includes backgrounds from interactions with correlated nucleons (2p2h) and single pion production ($1\pi$) with subsequent absorption.
The statistics considered here correspond to $4\times 10^{21}$~protons on target (POT) in the two tons of active target with the T2K flux, expected to be gathered with the near detector upgrade before the end of T2K (assuming data taking predominantly in neutrino mode), or during six years of \hk data taking at a beam power of \SI{1.3}{\mega\watt}~\cite{Hyper-Kamiokande:2018ofw}.
It can be seen how the bulk of the distributions, dominated by CCQE interactions, undergoes a shift between the models, with statistical sensitivity to shifts on the few-MeV level, where the distributions are the most sensitive in the rising and falling edges around $\pm \SI{150}{\mega\eVperc}$.
It is encouraging to see that, even after consideration of FSI, the shifts in the missing neutrino energy are well correlated with the changes in the underlying removal energy. More crucially, the shifted removal energy manifests as an almost direct shift to an observable, $\dplvis$, even when considering effects from detector smearing, background components and FSI.

\begin{figure}[tb]
    \centering
    \includegraphics[width=\linewidth]{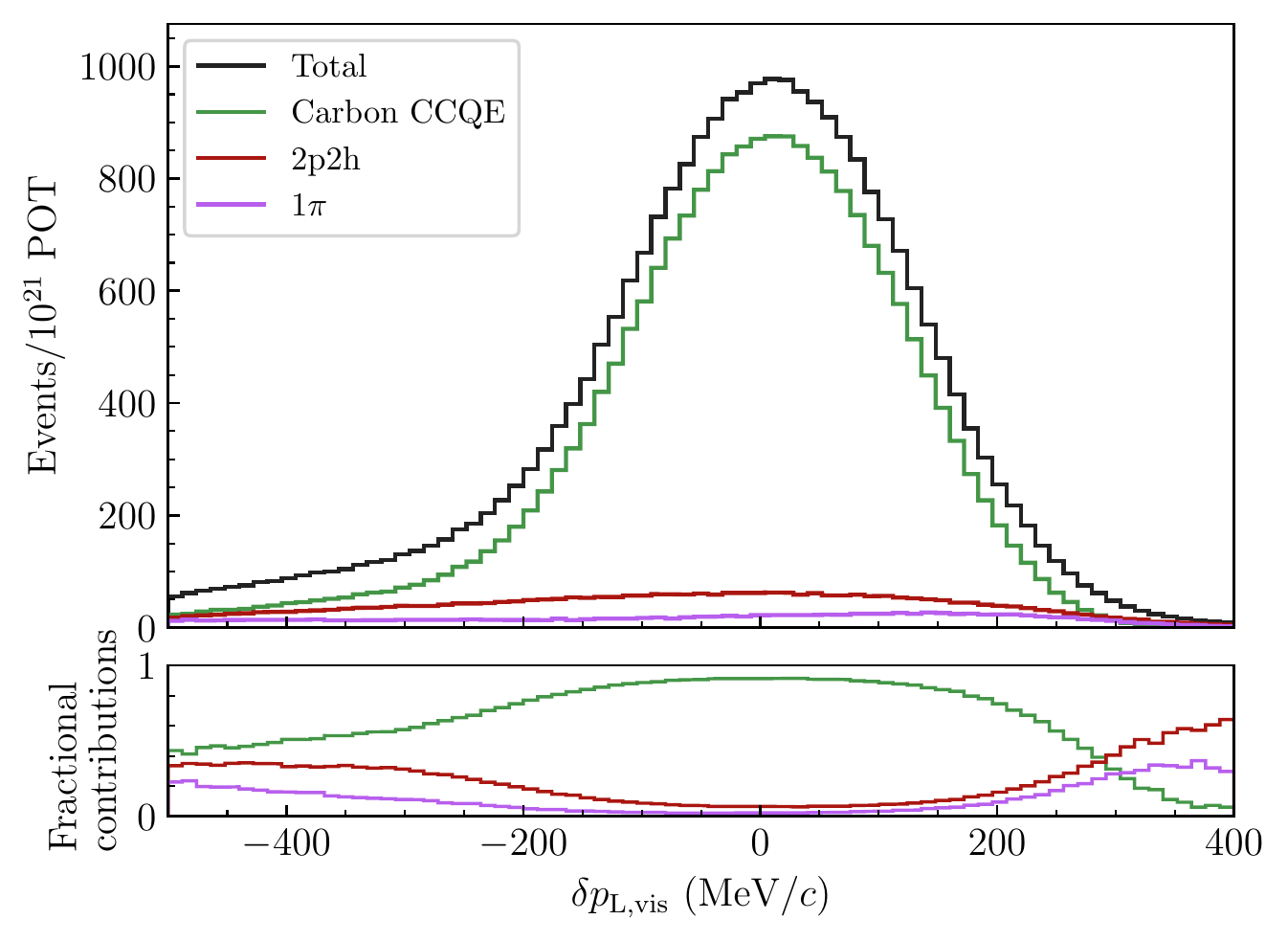}
    \includegraphics[width=\linewidth]{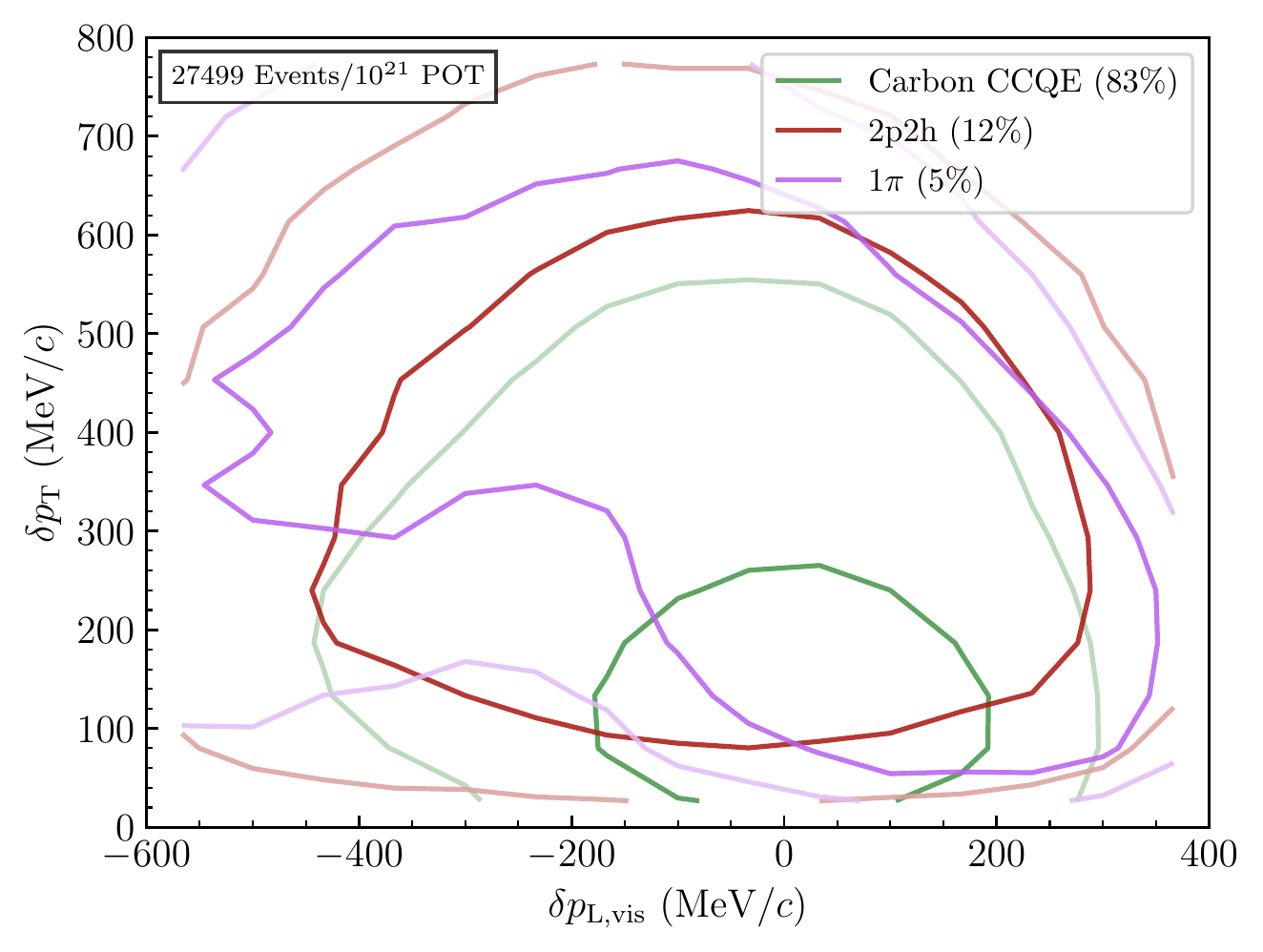}
    \caption{The distribution of $\dplvis$ (top) and $\dplvis$ against $\dpt$ (bottom), where detector smearing is applied to CC0$\pi$ events with one muon and one proton detected in the final state. In the two-dimensional distribution, the two contours for each distribution enclose \SI{68}{\percent} and \SI{95}{\percent} of the events, respectively, with the overall purities indicated in the legend.
    }
    \label{fig:nu_dpLvis_dpT}
\end{figure}

Considering the role of NEUT's FSI model, as well as the non-CCQE backgrounds including 2p2h and pion production events, the distribution of $\dplvis$ is broad, extending to a negative tail. This is due to undetected contributions to the longitudinal momentum, where the undetected kinetic energy contributes less in magnitude to $\dplvis$. This effect can be observed in \cref{fig:dplvis_sf_models_4e21pot}, and is explicitly shown in \cref{fig:nu_dpLvis_dpT}, where the $\dplvis$ distribution in a CCQE-like selection is shown split by interaction mode.
Further, the shapes of the two-dimensional distributions with $\dpt$ are shown.
While the CCQE interactions are less separated in $\dplvis$ compared to $\dpt$, it can be seen how the combination of both variables may nonetheless yield an improved separation of the interaction modes.

We compared the shift in $\dplvis$ to $\dpty$, 
defined as the transverse momentum imbalance along the direction of the leptonic transverse momentum, suggested in \ccite{Cai:2020nbe} 
to be sensitive to the removal energy. However, we found the CCQE component to shift by less than \SI{0.5}{\mega\eVperc} between the models before detector effects, offering a much reduced sensitivity compared to the \SI{10}{\mega\eVperc} shift in $\dplvis$.
Similar results were found for the other STVs.

\begin{figure}[tb]
    \centering
    \includegraphics[width=\linewidth]{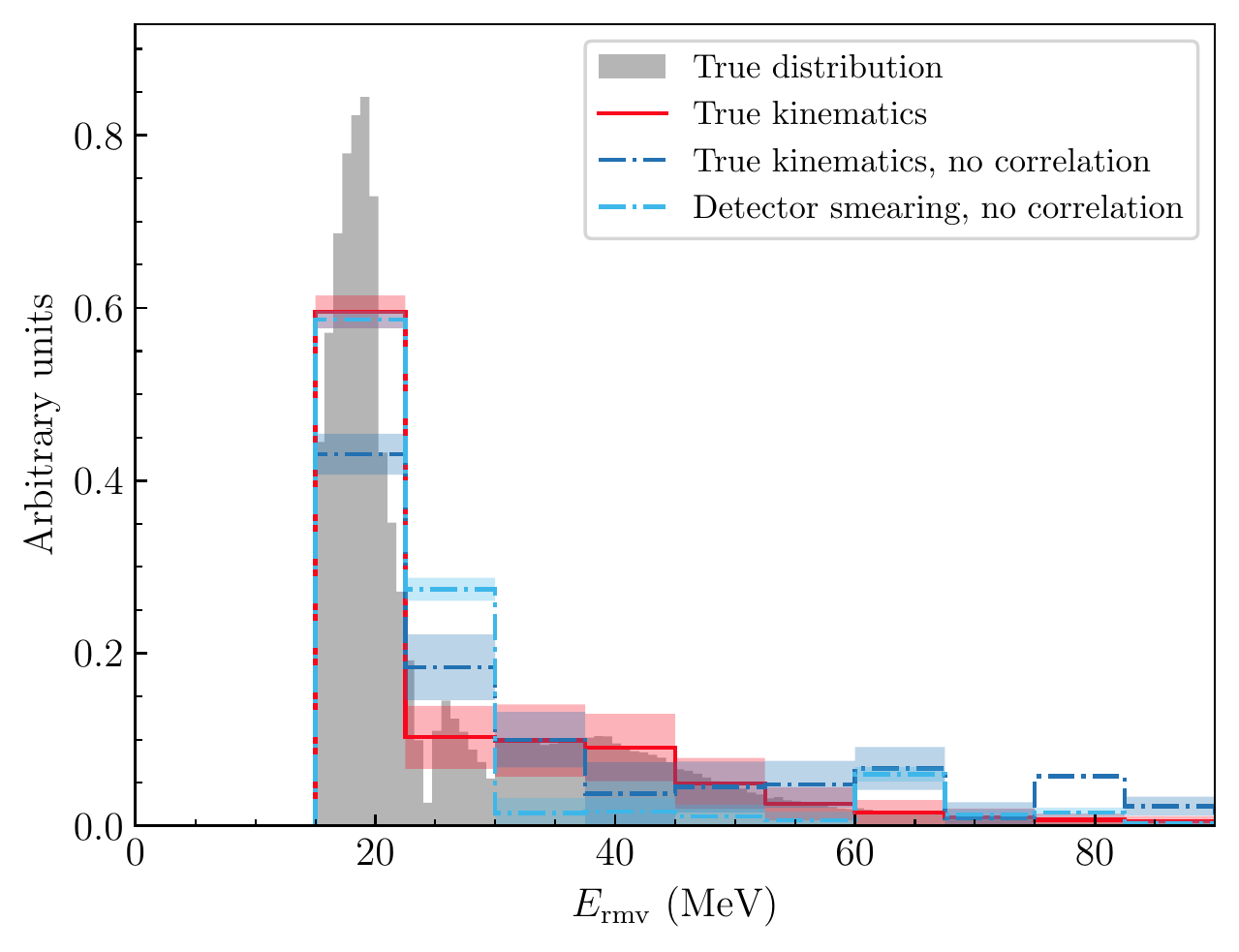}
    \caption{The results of the $\ermv$ likelihood fit to Asimov datasets using different templates.
    The uncertainties on the fit parameters are indicated by the shaded regions, and the filled distribution shows the underlying true distribution from the SF model in NEUT.
    }
    \label{fig:Ermv_fit}
\end{figure}

Considering an analysis beyond this overall shift, we demonstrate that $\dplvis$ has the potential %
to constrain the shape of the removal energy distribution by performing a simple fit.
We use a selection of pure CCQE interactions in the NEUT~SF model and reconstruct the removal energy distribution from a simple template fit in $\dplvis$. Ten free parameters describe the contributions to the removal energy in a range from \SIrange{15}{90}{\mega\eV} in uniform steps, where template distributions in $\dplvis$ are generated for these different $\ermv$ intervals. This assumes the initial momentum distribution $\pnl$ is known, which may be obtained from a measurement of the transverse momentum imbalance as mentioned above. A more detailed description of the methodology can be found in Appendix~\ref{sect:appendix_fit_methods}.
We perform a fit to the Asimov dataset, generated with a uniform weighting of the templates, using the true kinematics to reconstruct $\dplvis$. 
It is not only the overall removal energy distribution which impacts the $\dplvis$ distribution, but also its correlation with the initial nucleon momentum, as described by the spectral function.
To explore this effect, we then fit to the same dataset but using different templates which assume no correlation in the spectral function, and further with detector smearing applied to the templates and dataset. 
Note that here and unlike shown above in~\cref{fig:dplvis_sf_models_4e21pot}, no detector efficiencies or background events are included. 
The fit results are shown in \cref{fig:Ermv_fit}.
Using the true kinematics with the correct underlying spectral function, the fit reconstructs the shape of the $\ermv$ distribution.
When assuming no correlation and adding detector smearing in addition, the reconstruction is worsened, yet the broad features remain.
In these cases the fit agreement is worse, meaning that the no-correlation assumption does not permit the correct $\dplvis$ shape to be reconstructed. This also explains the smaller fit errors, related to the more off-diagonal correlation matrix, shown in Appendix~\ref{sect:appendix_fit_methods}.
This effect shows how the $\dplvis$ distribution and its comparison with the transverse momentum imbalance may not only constrain the overall removal energy shape, but also its correlation with the initial nucleon momentum, which is stronger in the relativistic (RFG) and local (LFG) Fermi gas models for instance.
Such a correlation could be modeled with additional parameters in a more complex fit.
Further, a more advanced analysis would proceed with a joint fit of $\dplvis$ and $\dpt$, including a modeling of the second-order differences between $\pnl$ and the transverse imbalance.
We leave a more quantitative sensitivity study proceeding along these lines for future work. We additionally remind the reader that this fit was performed using a model that, whilst considering FSI through the use of a cascade model, does not directly simulate a nuclear potential and that a full analysis would offer a constraint instead on the collective impact of this and the removal energy.

\subsection{High-purity hydrogen sample in antineutrino interactions}

In antineutrino interactions on a plastic scintillator (C$_8$H$_8$) detector,
interactions can occur on the free protons making up the hydrogen nuclei, which are free of nuclear effects.
In \ccite{Munteanu:2019llq}, the use of the transverse momentum imbalance $\dpt$ to select a sample enriched with interactions on hydrogen has been investigated for a segmented plastic scintillator detector such as in \ccite{Blondel:2017orl},
which can reconstruct neutrons from their secondary interactions in the detector using the time of flight (ToF) method.
Thanks to its reduced bias from nuclear effects, such a sample shows an improved resolution on the neutrino energy, delivering an enhanced constraint on the neutrino flux. The reduction in the flux normalization uncertainty for the upgraded ND280 detector has been quantified in \ccite{Dolan:2022sut}. As explored therein, the extraction of neutrino interactions on a hydrogen sample can also be used to constrain nucleon form factors in a way that is free from degeneracies with nuclear effects.
In this section, the same analysis strategy is used as in \ccite{Munteanu:2019llq}, where the neutron detection efficiency and resolutions that were obtained from an external simulation are applied to simulated neutrino interaction events, generated using NEUT with the T2K flux.
We add the longitudinal momentum imbalance $\dplvis$ to the analysis, in this case computed as $\dplvis = p_{n,\mathrm{L}} + p_{\mu,\mathrm{L}} - (E_n + E_\mu - m_p)/c$ using the kinematics of the final state neutron and muon in the CCQE-like sample, after applying detector effects.
Note that here, no \quotes{lever arm} cut is applied, \ie no minimum distance to the secondary neutron interaction cluster is required.

\begin{figure}[htb]
    \centering
    \includegraphics[width=\linewidth]{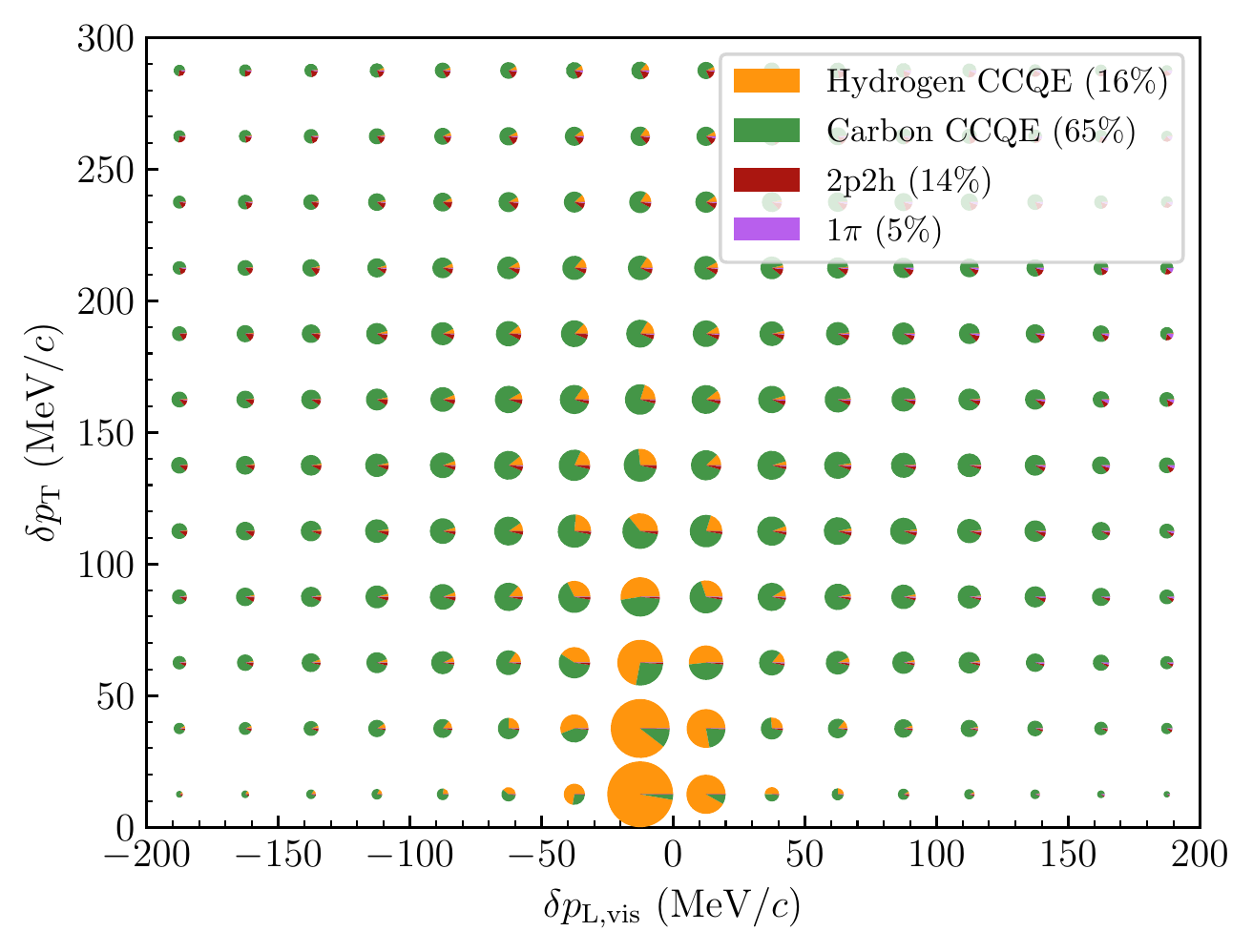}
    \caption{The distribution of antineutrino interactions in $\dpt$ and $\dplvis$ with 
    detector smearing applied, split by interaction mode and target. The area of each circle is proportional to the number of events within the corresponding bin, and the overall purities before any cuts are shown in the legend.}
    \label{fig:anu_dplvis_dpt_pie}
\end{figure}

CCQE interactions without FSI are already relatively well separated from the other events by $\dpt$, as shown for instance in \ccite{Munteanu:2019llq,Dolan:2022sut}.
For interactions on hydrogen, just as $\dpt$, $\dplvis$ is zero before detector smearing effects. When isolating a hydrogen sample, $\dplvis$ can thus be used to further reject interactions on carbon passing the $\dpt$ cut, being especially useful for events where the initial nucleon momentum was oriented along the longitudinal direction.
The two-dimensional distribution in $\dpt$ and $\dplvis$ of antineutrino interactions with 
detector smearing are shown in \cref{fig:anu_dplvis_dpt_pie}, split by interaction mode and target. The addition of $\dplvis$ can be seen to provide an additional separation of interactions on hydrogen, which show a high purity for $\dpt$ and $\dplvis$ both close to zero.
It should be noted that in the absence of detector smearing effects, $\dpt$ perfectly separates the interactions on hydrogen (as shown in \ccite{Munteanu:2019llq}); the additional degree of freedom from the longitudinal direction is however beneficial when such effects are considered, as seen here.

\begin{figure}[htb]
    \centering
    \includegraphics[width=0.98\linewidth]{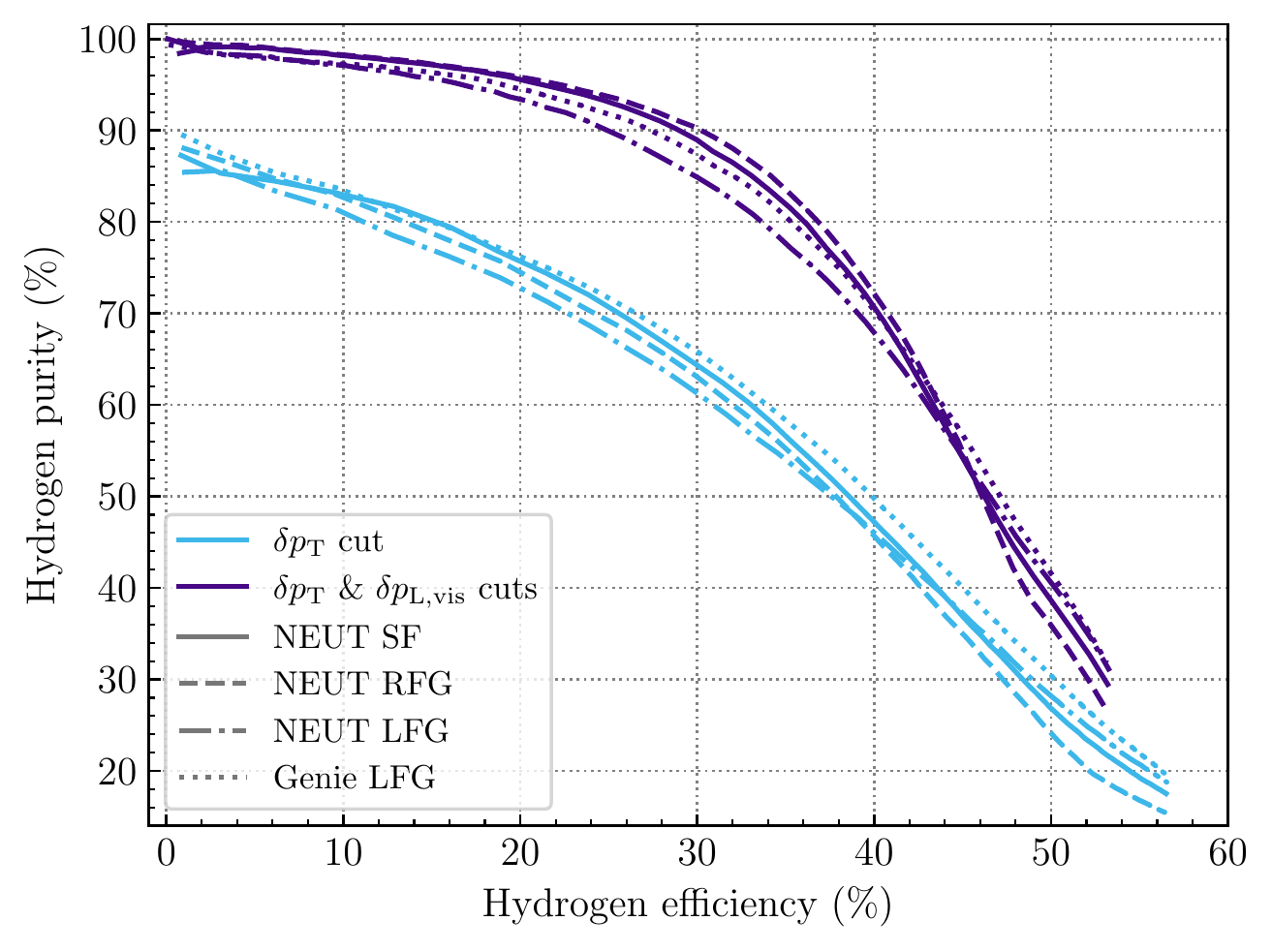}
    \caption{The purity vs.\ efficiency of antineutrino interactions on hydrogen for cuts on $\dpt$ and combined cuts in $\dpt$ and $\dplvis$. Different models for the nuclear initial state of carbon are shown.}
    \label{fig:hyd_pur_eff_dpLvis_models}
\end{figure}

For a given efficiency of interactions on hydrogen, the combination of cuts in $\dpt$ and $\dplvis$ which maximizes the hydrogen purity is selected.
Due to the non-linear relation between energy and momentum, the effect of detector smearing on $\dplvis$ is asymmetric, as can be seen in \cref{fig:anu_dplvis_dpt_pie}, where the hydrogen events are smeared away from zero with a bias towards negative values. 
As such, the cuts on $\dplvis$ are centered around \SI{-6}{\mega\eVperc}.
The results on the purity vs.\ efficiency of hydrogen events are shown in \cref{fig:hyd_pur_eff_dpLvis_models}, both for cuts on $\dpt$ alone (as previously performed in \ccite{Munteanu:2019llq}) and with $\dplvis$ in addition as described above.
Different models for the nuclear initial state of the carbon component are compared, including RFG and LFG~models from NEUT and GENIE~\cite{Andreopoulos:2009rq}. 
There is a small systematic spread between the models, which does not increase substantially when adding the longitudinal information.
Overall, the use of the longitudinal kinematic imbalance brings a drastic increase in the hydrogen purity at a given efficiency, and vice versa. For instance, at a hydrogen efficiency of \SI{20}{\percent}, the purity is increased from \SI{77}{\percent} to \SI{96}{\percent} for the SF model, reducing the relative background by a factor of more than five. 
As a reference, around \num{4000}~neutrino-hydrogen interactions per \SI{e21}{\pot} are expected in the two~ton \sfgd active mass before considering detector efficiencies. At a hydrogen purity of \SI{90}{\percent} and with the efficiencies considered here, around \num{27000}~events would be obtained throughout \num{10}~years of \hk data taking.

\section{Discussion}

The visible longitudinal momentum imbalance $\dplvis$ introduced above is shown to offer sensitivity to the collective nuclear effects %
in neutrino interactions which bias neutrino energy reconstruction, to a greater extent than STVs.
While it is dominated by the initial state nucleon's longitudinal momentum, an overall shift in the distribution has been shown to be sensitive to the average removal energy, which can be a crucial observable to discriminate between the different nuclear models available in literature, as well as reduce the bias on the neutrino energy and thus the neutrino oscillation parameters. We have noted that in more sophisticated models that include the role of a nuclear potential we expect the shift seen in $\dplvis$ to be sensitive to the combined effect of this and the removal energy (our ``missing neutrino energy''). Whilst one cannot easily be separated from the other, it is the combined effect of both that drives biases in neutrino energy reconstruction and so a measurement remains a powerful tool of constraining key systematic uncertainties in measurements of neutrino oscillations.
We have additionally shown that $\dplvis$ has the potential to provide shape information on the nuclear removal energy distribution as well as its complementarity with transverse kinematic imbalances in isolating contributions from distinct interaction channels in CC0$\pi$ cross-section measurements.
Fine-granularity detectors with a resolution on the millimeter scale, including liquid argon time projection chambers and 3D~scintillator detectors, may measure the kinematic imbalances with an increased precision, thereby obtaining a further sensitivity to the shape information.

The method presented here is not unique in constraining the missing neutrino energy: %
As seen from \cref{eq:ccqe_ia_e}, the average removal energy also statistically shifts the visible final state energy in relation to the true neutrino energy distribution. 
Detecting the shift in $\evis$ however requires a relatively precise prediction of the incoming neutrino flux, with any bias propagated to the prediction of the removal energy.
Yet as shown above, $\dplvis$ has a minimal dependence on the neutrino energy and thus the flux prediction.
Naturally, the method of using $\dplvis$ introduces its own systematic uncertainties due to the detailed second-order nuclear effects, including Pauli blocking and final state interactions. 
An analysis of neutrino interaction data could then proceed with a multidimensional fit to $\evis$, $\dplvis$ and $\dpt$, reducing the overall uncertainty on the missing neutrino energy. %
As mentioned above, a full systematic study will be required in order to more quantitatively study the full sensitivity.
Note that in our analysis of missing neutrino energy sensitivity, %
we focused on neutrino interactions, producing a proton in the final state. In principle, this method is also applicable to antineutrino interactions by extracting the shift in $\dplvis$ of the bulk of CCQE interactions on carbon. However, obtaining constraints in this case would be further made difficult by the increased systematic uncertainties associated with the reconstruction of the final state neutrons.

In antineutrino pion-less interactions, the use of the longitudinal kinematic imbalance next to the transverse imbalance was shown to provide a high-purity sample of neutrino interactions on a hydrogen nuclear target from an initial sample of interactions on hydrocarbon (the target material in scintillator detectors).
In addition to delivering a strong constraint on the flux for a neutrino oscillation experiment, such a sample can be employed to measure neutrino-nucleus interactions minimally biased by nuclear effects, as described by the axial vector form factor $\fa$~\cite{Dolan:2022sut,Cai:2019jzk}. The \minerva collaboration has recently performed the first measurement of $\fa$ in antineutrino interactions with hydrogen, with a signal purity and efficiency of around \SI{30}{\percent} and \SI{10}{\percent} respectively~\cite{Cai:2023mav}. 
The method presented here can be applied to any detector technology with a hydrogen content and the capability to reconstruct the outgoing neutron momentum.
Without sufficient timing resolution and 3D~granularity, the \minerva detector only measures the direction of outgoing neutrons propagating in the forward direction, such that this method is not applicable.
A 3D segmented plastic scintillator such as the 
\sfgd 
considered in this work is instead well suited due to its ability to reconstruct neutron momenta using the ToF method.
The same detector technology has been investigated for a potential near detector in the future DUNE experiment, with a mass of 10~tons~\cite{Gwon:2023nda}.

A flux-constraining method similar to the antineutrino one detailed above may be achieved with deuterated carbon scintillators, while providing quasi-free nucleon data, where the neutrino can undergo a CCQE interaction with the neutron in deuteron, minimally biased by nuclear effects.
We performed a similar analysis to the antineutrino case by simulating neutrino interactions on deuterated plastic, finding for instance that at a deuteron efficiency of \SI{10}{\percent}, a purity of \SI{78}{\percent} (\SI{64}{\percent}) can be achieved  for pure (half) deuterated plastic.
However, such a technology remains to be studied for a neutrino detector, in particular from a point of view of feasibility and cost.

\section{Conclusion}

In conclusion, we have introduced a powerful new observable, the visible longitudinal momentum imbalance ($\dplvis$), for accelerator neutrino oscillation experiments. 
Combined with the observed transverse momentum imbalance ($\dpt$), it can deliver improved constraints on fundamental nuclear uncertainties, not directly accessible to current experiments, particularly those which bias the reconstruction of the neutrino energy.
Further, it can allow for a high-purity selection of antineutrino interactions on hydrogen, which would lead to the precise measurement of the nuclear axial vector form factor as well as of the antineutrino flux.

\begin{acknowledgments}

DS was supported by the Swiss National Science Foundation Eccellenza grant (SNSF PCEFP2\_203261), Switzerland.

\end{acknowledgments}

\appendix

\section{Unexplained behavior in the SF model}
\label{sect:appendix_sf_shift}

\begin{figure*}[htb]
    \centering
    \includegraphics[width=0.49\linewidth]{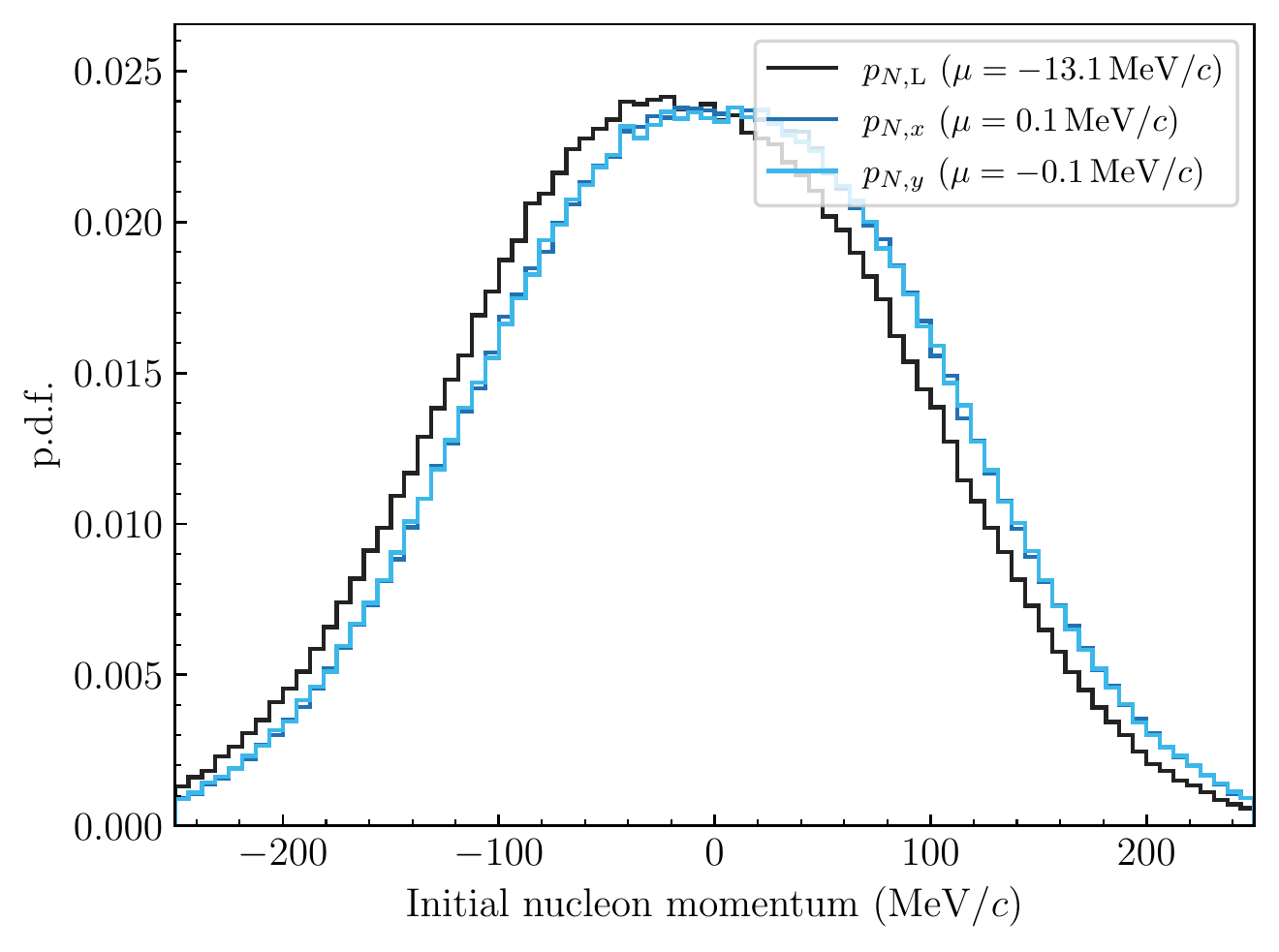}
    \includegraphics[width=0.49\linewidth]{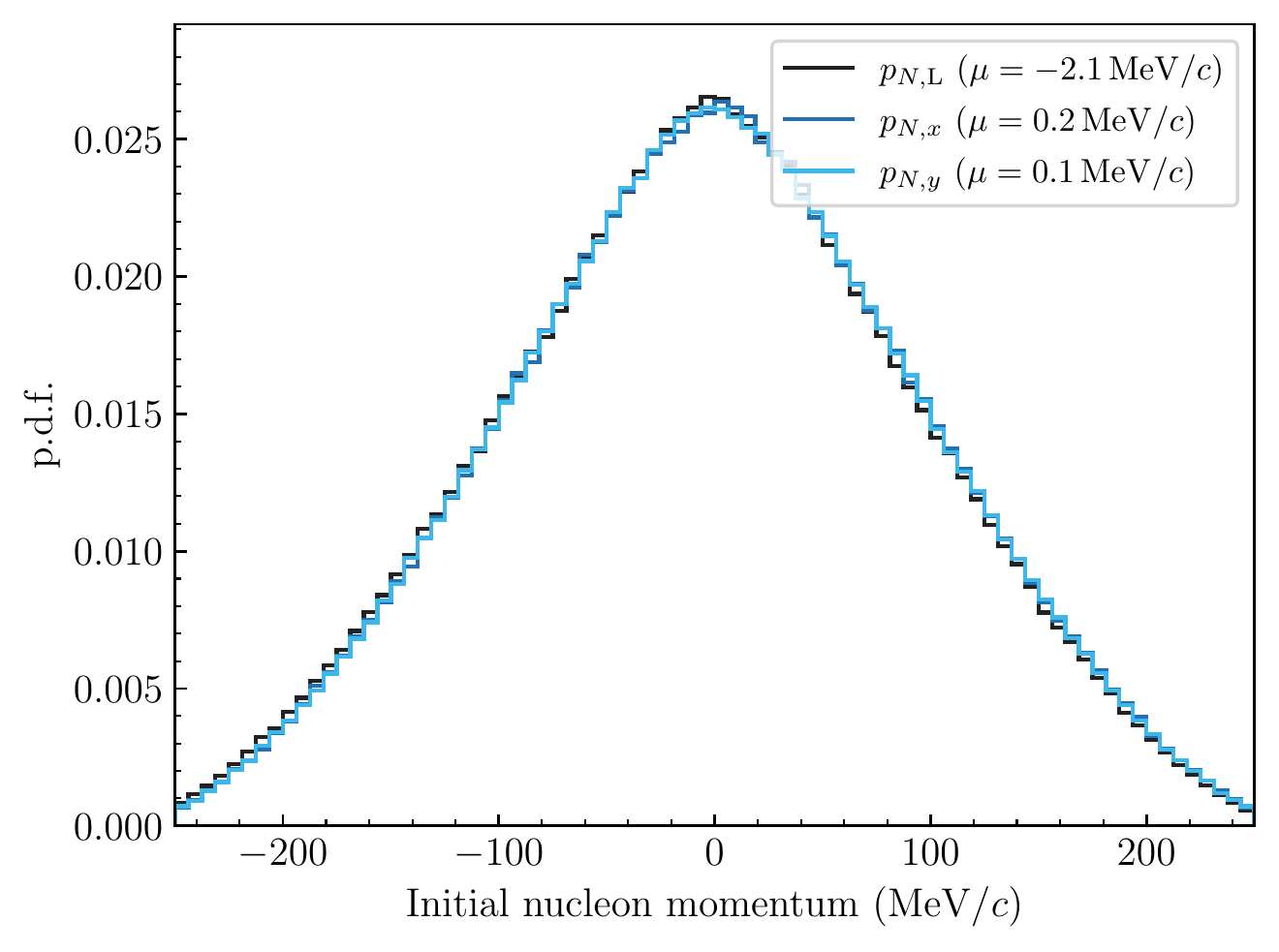}
    \caption{The distributions of initial nucleon momenta $\pni$ in CCQE interactions without FSI with one muon and one proton in the final state, for the NEUT~SF (left) and LFG (right) models.
    }
    \label{fig:pnL}
\end{figure*}

We detail here the unexplained behavior that was observed in neutrino event generator implementations of the SF model.
While the values given here correspond to the output from NEUT, the same effect was observed in the NuWro implementation.
The initial proton longitudinal momentum ($\pnl$) distribution in CCQE interactions
appears to undergo an overall shift from a symmetric (isotropic) one to a mean of around \SI{-13}{\mega\eVperc}.
Due to the size and uniformity of this shift, it is inconsistent with the second-order effects such as polarization and Pauli blocking which were mentioned in the main text. Furthermore, this behavior is not observed in the RFG or LFG models, where the $\pnl$ distributions only undergo a shift of \SIrange{2}{3}{\mega\eVperc}.
The distribution of $\pnl$ is shown in \cref{fig:pnL} for the SF and LFG models alongside the transverse momentum imbalances.

With this behavior persisting in a one-proton final state sample with FSI disabled, this rules out FSI and short-range correlations (SRCs) in the SF model as a possible cause, with the latter producing a multi-nucleon final state. 
In a sample with Pauli blocking disabled in addition to FSI and a high neutrino energy of \SI{20}{\giga\eV} such that the polarization effect is minimized, a shift of around \SI{-10}{\mega\eVperc} remains.
This shift is propagated to the $\dplvis$ distributions that are shown %
in the main text, where %
the relation $\dplvis = \pnl + \ermv$ still holds in the generator output.

\section{Fit methodology and correlation matrices}
\label{sect:appendix_fit_methods}
We use a pure CCQE selection with one proton and one muon in the final state in all cases, \ie without background, generated with the NEUT~SF model.
Two Asimov datasets are generated from a uniform weighting of the templates scaled to \num{300000}~events, with and without detector smearing applied.
In the case of no correlations, both without and with detector smearing, we fit to the regular Asimov datasets, but the templates are re-weighted such that there is no correlation between the true $\pnl$ and the removal energy (aside from the small contribution from the kinetic energy of the nuclear remnant, which is correlated with the initial nucleon momentum), where the overall $\pnl$ distribution matches that of the SF model.
The best fit removal energy contribution is obtained by finding the ten parameters which maximize the binned multinomial likelihood of the reconstructed $\dplvis$ distribution with respect to the dataset in consideration. 
Both are binned in widths of \SI{3.5}{\mega\eVperc} as this was found to better preserve the shape information compared to larger bin sizes.
In each fit, the template weights are randomly initialized between zero and two. This procedure is repeated 1000~times, from which the best fit is selected.
To avoid issues with parameter boundaries, the parameters are allowed to have negative values, but their absolute value is used to compute the reconstructed $\dplvis$ distribution.

\begin{figure*}[htb]
    \centering
    \includegraphics[width=0.49\linewidth]{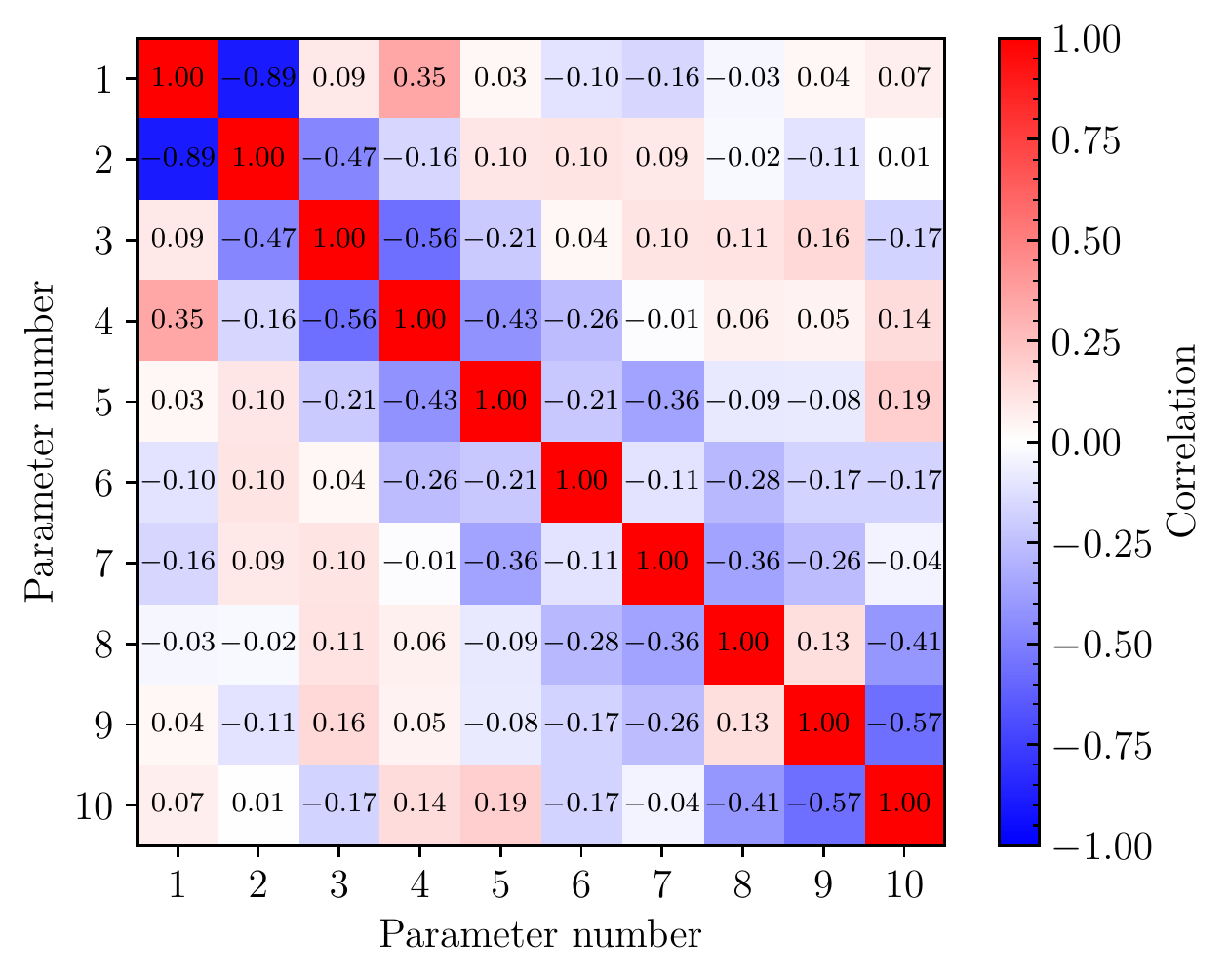}
    \includegraphics[width=0.49\linewidth]{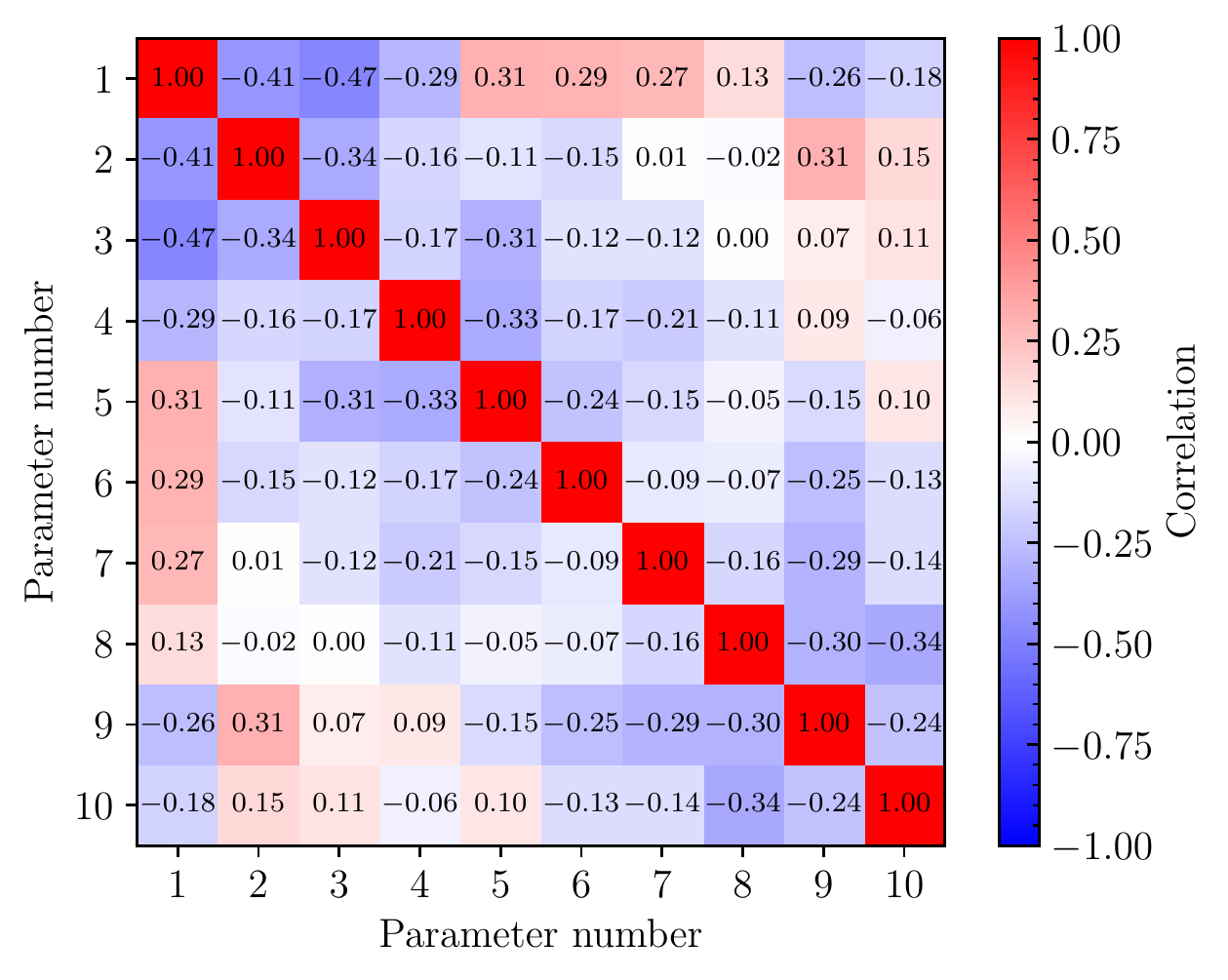}
    \caption{The post-fit correlation matrices from the removal energy fit to the Asimov dataset with the true kinematics (left) and the smeared dataset using templates with no correlations (right).
    }
    \label{fig:Ermv_fit_corr_matrices}
\end{figure*}

The correlation matrices from the fits are shown in \cref{fig:Ermv_fit_corr_matrices}. The fit to the Asimov dataset shows stronger (anti)correlations between parameters describing adjacent bins in $\ermv$, while the case with the altered no-correlation templates shows more uniform correlations, in particular between parameters describing non-adjacent bins.

\onecolumngrid

\twocolumngrid

 \newcommand{\noop}[1]{}

\end{document}